\documentclass{elsart}

\usepackage{natbib}

\def\lsim{\mathrel{\raise.3ex\hbox{$<$\kern-.75em\lower1ex\hbox{$\sim$}}}} 
\def\gsim{\mathrel{\raise.3ex\hbox{$>$\kern-.75em\lower1ex\hbox{$\sim$}}}} 
\usepackage{graphicx}

\usepackage{amssymb}

\begin{document}

\begin{frontmatter}



\title{Probing Supersymmetric Parameters With Astrophysical Observations}


\author{Dan Hooper}

\address{Fermi National Accelerator Laboratory, Theoretical Astrophysics Group, Batavia, IL  60510}

\begin{abstract}

A wide range of techniques have been developed to search for particle dark matter, including direct detection, indirect detection, and collider searches. The prospects for the detection of neutralino dark matter is quite promising for each of these three very different methods. Looking ahead to a time in which these techniques have successfully detected neutralino dark matter, we explore the ability of these observations to determine the parameters of supersymmetry. In particular, we focus on the ability of direct and indirect detection techniques to measure the parameters $\mu$ and $m_A$. We find that $\mu$ can be much more tightly constrained if astrophysical measurements are considered than by LHC data alone. In supersymmetric models within the $A$-funnel region of parameter space, we find that astrophysical measurements can determine $m_A$ to roughly $\pm100$ GeV precision. 


\end{abstract}


\end{frontmatter}

\section{SUSY Parameters From Direct and Indirect Detection}
\label{}

If low scale supersymmetry exists in nature, it will likely be discovered in the next few years at the Large Hadron Collider (LHC), or perhaps even earlier at the Tevatron. In many supersymmetric scenarios, the LHC can identity the presence of the lightest neutralino in the form of missing energy in the cascade decays of squarks and/or gluinos. At approximately the same time, direct and indirect searches for dark matter will be reaching the level of sensitivity needed to discover neutralinos in a wide range of supersymmetric models.

Collider and astrophysical experiments tell us very different things about the nature of supersymmetry. The LHC is likely to reveal the approximate masses of a number of superpartners, including squarks, gluinos, the lightest neutralino, and in some cases sleptons and heavier neutralinos. Other properties of the supersymmetric model will remain unconstrained by the LHC, however. In particular, the composition of the lightest neutralino (the mixture of bino, wino, and higgsino components), and therefore its couplings, will be very difficult to deduce at a hadron collider. In contrast, the cross sections relevant to astrophysical dark matter experiments depend critically on the composition of the lightest neutralino.

\begin{figure}[t]
\hspace{-1.5cm}
\includegraphics[width=2.2in,angle=-90]{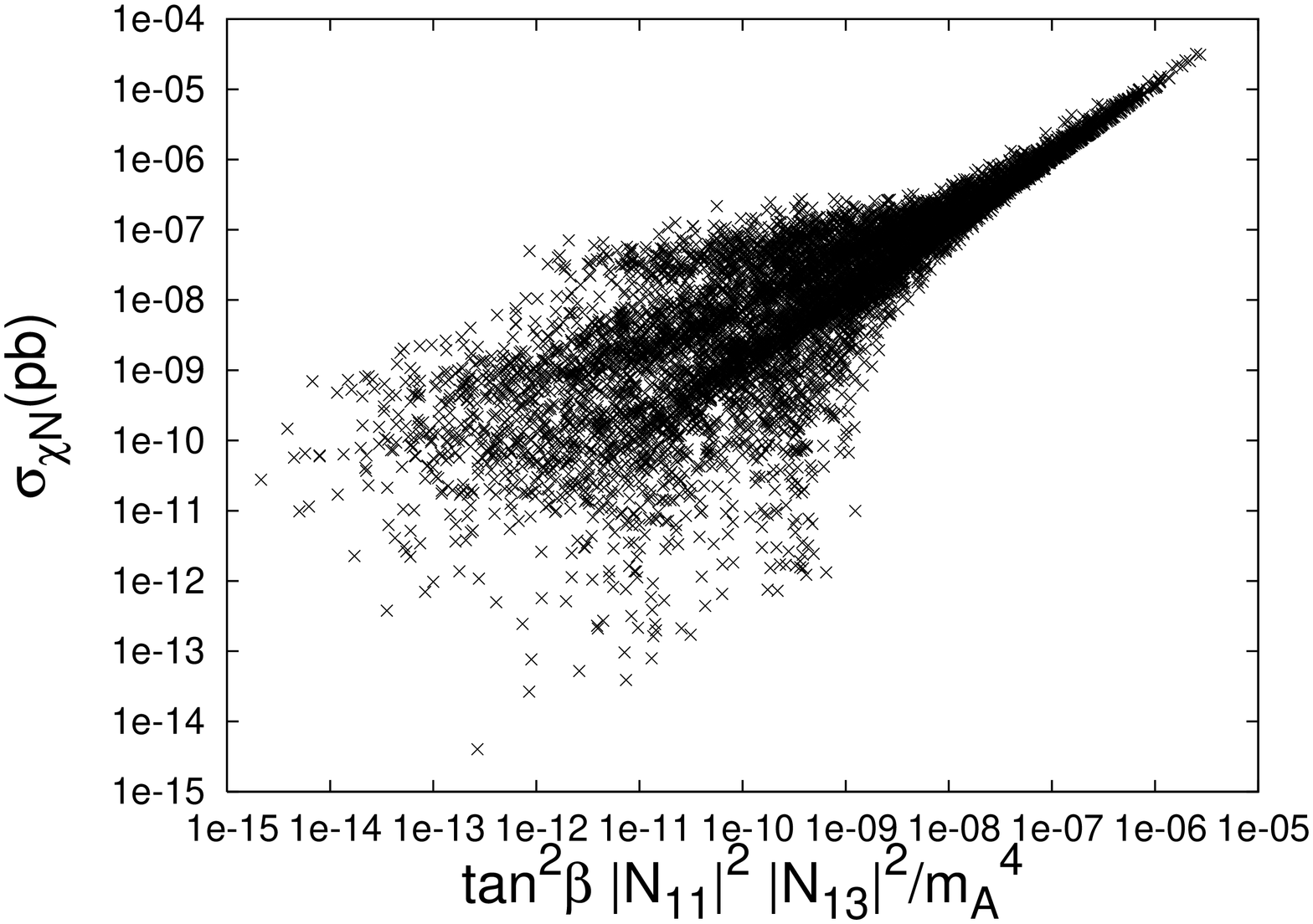}
\includegraphics[width=2.2in,angle=-90]{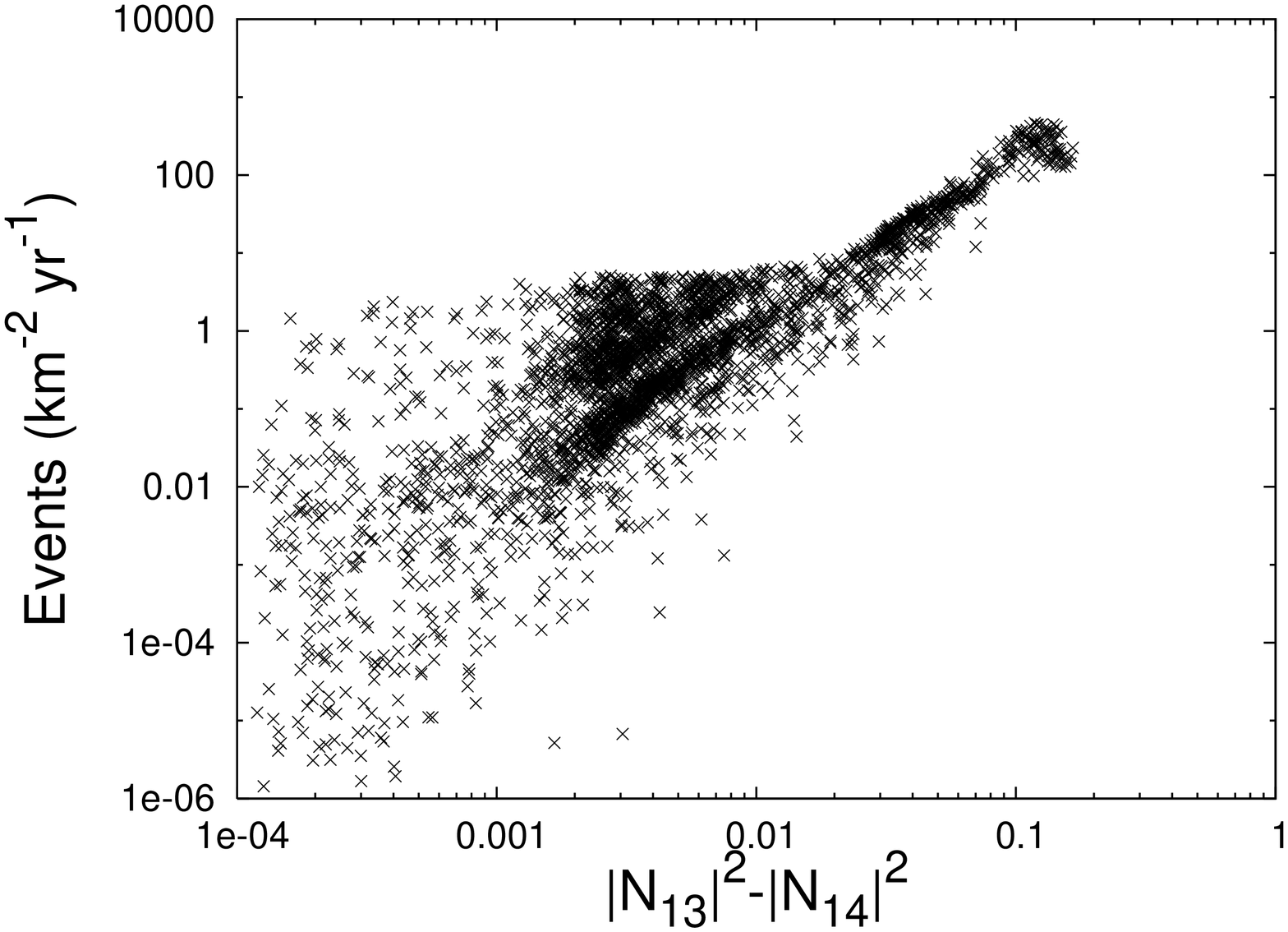}
\caption{Left: The relationship between the quantity  $|N_{11}|^2 |N_{13}|^2 \tan^2 \beta/m^4_A$ and the spin-independent neutralino-nucleon elastic scattering cross section. Right: The rate of neutrinos detected in a kilometer scale neutrino telescope, such as IceCube, from neutralino annihilations in the Sun as a function of the quantity $|N_{13}|^2-|N_{14}|^2$. A constraint 100 times more stringer than the current CDMS bound has been applied (in the right frame) in anticipation of increased sensitivity from direct detection experiments in the coming few years.}
\label{directneutrino}
\end{figure}

In Fig.~\ref{directneutrino}, we show how the rates in astrophysical dark matter experiments depend on the composition of the lightest neutralino. In the left frame, we compare the scalar neutralino-nucleon elastic scattering cross section (which determines the rate in direct dark matter experiments) to the quantity:  $\tan^2 \beta |N_{11}|^2 |N_{13}|^2/m^4_A$. Here, $m_A$ is the mass of the CP-odd Higgs boson in the MSSM. The $|N|^2$'s are defined by the following: $\chi^0_1 = N_{11}\tilde{B}     +N_{12} \tilde{W}^3
          +N_{13}\tilde{H}_1 +N_{14} \tilde{H}_2$. The correlation shown in the left frame of Fig.~\ref{directneutrino} is quite tight for models with large cross sections, which are produced through diagrams which exchange a heavy Higgs boson  coupling to $b$ and $s$ quarks (\cite{carena}).

Neutralinos scattering with the Sun can become gravitationally bound and accumulate in the Sun's core, eventually annihilating to produce high-energy neutrinos. The rate of those neutrinos being observed at a kilometer-scale neutrino telescope such as IceCube or KM3 is shown in the right frame of Fig.~\ref{directneutrino} as a function of the quantity $|N_{13}|^2-|N_{14}|^2$. This correlation comes from the $\chi^0$-$\chi^0$-$Z$ coupling which largely determines the capture rate in the Sun in nearly all observable models (\cite{halzen}).

\section{An Example of Constraining $\mu$}
\label{mu}

We will next consider a benchmark model in order to study how direct and indirect detection information can aid in determining the composition of the lightest neutralino, and therefore the parameter $\mu$. We will adopt the following benchmark: $M_2=472.9$ GeV, $\mu= 619.2$ GeV, $\tan\beta=50.6$, $m_A=396.5$ GeV, and a common sfermion mass scale of 2130 GeV. This model generates a thermal relic abundance of $\Omega_{\chi^0} h^2=$0.098, which is consistent with the measurements of WMAP. The lightest neutralino in this model is 99.1\% bino, with a small higgsino admixture. Although IceCube/KM3 will not be sensitive to this model, the elastic scattering cross section of $9.6 \times 10^{-9}$ pb will be within the reach of next generation direct detection experiments.

\begin{figure}[!tbp]
\hspace{3.0cm}
\includegraphics[width=2.25in,angle=-90]{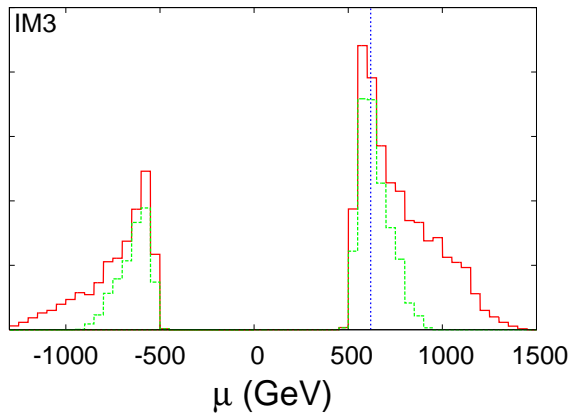}
\caption{The distribution of $\mu$ in the models found by our scan which match the features that would be observed at the LHC in our first benchmark model (upper curve). The low curve only contains those models which also match the features that would be observed by direct and indirect detection experiments.}
\label{mu}
\end{figure}

A number of supersymmetric particles will be within the reach of the LHC in this model. In addition to quarks, gluinos, and the lightest neutralino, $\tan \beta$ is large enough, and $m_A$ small enough, to be determined though the channel $H/A \rightarrow \tau^+ \tau^-$ at the LHC. In the upper curve of Fig.~\ref{mu}, we plot the distribution of SUSY models found with our scan that match the following constraints from LHC measurements: $m_{\tilde{q}}=2130$ GeV $\pm 30\%$, $m_{\chi^0}=236$ GeV $\pm 10\%$, $m_A=397\pm 3$ GeV, and $\tan \beta =51 \pm 15\%$, in addition to the relic density constraint. The lower curve on this figure, in addition to the LHC measurements, is constrained by direct and indirect detection measurements. Whereas the LHC alone can only constrain $1500 \gsim |\mu| \gsim 500$ GeV, the inclusion of astrophysical measurements can be used to tighten the upper bound considerably: $800 \gsim |\mu| \gsim 500$ GeV.

\section{An example of measuring $m_A$}

In models with low to moderate $\tan\beta$ or somewhat heavy $m_A$, heavy Higgs bosons will be beyond the reach of the LHC. In models in which neutralinos annihilate primarily through $A$-exchange, however, known as the $A$-funnel region, the value of $m_A$ can potentially be determined through dark matter observables.

\begin{figure}[!tbp]
\hspace{3.0cm}
\includegraphics[width=2.25in,angle=-90]{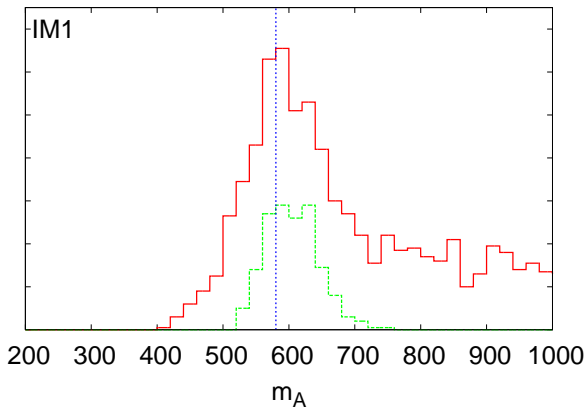}
\\
\caption{The distribution of $m_A$ in the models found by our scan which match the features that would be observed at the LHC in our second benchmark model (upper curve). The low curve only contains those models which also match the features that would be observed by direct and indirect detection experiments.}
\label{ma}
\end{figure}

Adopting a second benchmark model ($M_2=550.8$ GeV, $\mu= 1318$ GeV, $\tan\beta=6.8$, $m_A=580.2$ GeV, and a common sfermion mass scale of 2239 GeV), we plot in Fig.~\ref{ma} the ability of astrophysical measurements to determine the value of $m_A$. In this case, we find a determination of $m_A \approx 620 \pm 100$ GeV.  Since the sfermions are all heavy in this model, to avoid being overproduced in the early universe, either the neutralino must contain a substantial higgsino component, or be annihilating near the $A$-resonance. Direct detection experiments will find in this model $\sigma_{\chi N} \lsim 10^{-10}$ pb, constraining $|N_{13}|^2+|N_{14}|^2$ to be less than $\sim$0.1. Through this constraint, $2 m_{\chi^0}\approx m_A$ can be determined.

This talk was been based on work done with Andrew Taylor (\cite{hoopertaylor}).




\end{document}